\documentclass[a4paper]{article}

\usepackage{graphicx}

\begin{document}

\author{S. B. Goryachev \thanks{E-mail: goryachev@univ-lille1.fr}
\\Laboratoire de M\'{e}tallurgie Physique, URA CNRS n $¡$ 234,
\\B\^{a}t. C6, Universit\'{e} de Lille I, 59655 Villeneuve d'Ascq,
\\ France}

\date{14 October 1997}

\title{Kinetics, Hydrodynamics and Stochastodynamics of Cellular Structure 
Coarsening}

\maketitle

\begin{abstract} 
For the first time the phenomenon of cellular structure 
coarsening are consistently analysed from the positions of kinetic, 
hydrodynamic and stochastodynamic theories of nonequilibrium statistical 
systems.  Thereby micro-, meso- and macroscopic levels of approach are 
distinguished.  At the microscopic level the cellular structure is 
describe by a probability distribution function in a phase space of cell 
coordinates and of cell sizes.  A kinetic equation for the function is 
written and a development to a hydrodynamic equation of a mesoscopic cell 
medium is realised.  It has the form of a diffusion-reaction equation with 
a negative "diffusion" coefficient and with a cell interface density 
playing the role of concentration.  Its analysis reveals a new effect of 
macroscopic patterning in the cell medium: existence of space-correlated 
stochastic fluctuations of the cell interface density. 
   \par PACS numbers: 02.50.-r, 05.20.-Dd, 47.53.+r, 82.20.-w 

\end{abstract}

\section{Introduction} 

There are many physical systems consisting of 
homogeneous domains separated by distinct boundaries, so called cellular 
structures: ordered domains in alloys, magnetic domains, gas bubbles in 
soap froth and in lipid monolayers, grains in polycrystals, etc.  
\cite{1,2,3,4}.  We have here a simple physical situation (Fig.  1
 \cite{2}),
\begin{figure}
\includegraphics{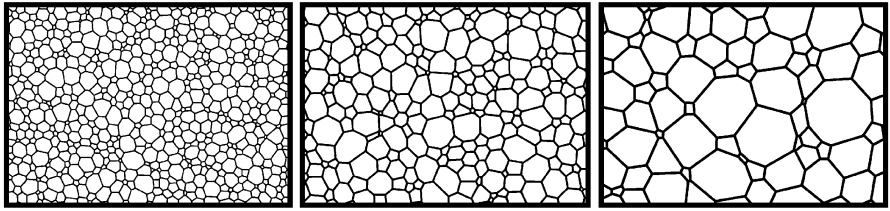}
\caption{Soap froth configurations at three different times during its 
evolution.}
\end{figure}
 when 
the system volume $V$ being the sum of volumes of all cells, the volume 
specific energy and the interface specific energy $E$ are constant. The 
internal energy of the system $\cal E$ is completely defined by the cell 
interface area $\Sigma$.  It is easy to see that for any two neighbouring 
cells the volume decrease of the smaller one (down to complete 
disappearance) and the simultaneous volume increase of the greater one 
diminish $\Sigma$.  This permanent topological possibility to decrease 
${\cal E}=E \Sigma$ makes the system intrinsically unstable.  The mean 
cell size increases during this statistical process (Fig.  1), named 
"coarsening" (large cells "eat up" small ones). The changes of $\cal E$ 
related to the elementary statistical event of cell disappearance are so 
large in comparison with the energy of thermofluctuations, that the 
coarsening cannot be described by ways of usual thermodynamics \cite{5}.  
Still this challenging statistical problem can be elegantly solved by 
methods of physical kinetics, as shown by Lifshitz, Slyozov and Hillert 
(LSH) \cite{6,7,8}, if we consider the cellular structure as some dynamic 
system.  To get a cell dynamics equation let us following LSH 
approximately consider each cell as an elastic sphere shell embedded in an 
elastic environment which represents the average effect of cell-cell 
interaction.  For a cell of radius $a$ we write an interface tension force 
as $F=-E(1/a-\kappa)$, where the term $1/a$ is related to the effect of 
interface surface tension of the cell itself and the term $\kappa$ - to 
the effect of interface surface tension of adjoining cells (the elastic 
environment).  The quantity $\kappa$ is nothing that a self-consistent 
mean of cell curvature throughout the system.  Believing that the cell 
interface velocity $w$ affected by $F$ is determined by viscous law 
$w=MF$, where $M$ is the cell interface mobility, we have \cite{6,7,8} 
\begin{equation} {\rm d}a/{\rm d}t=-D(1/a-\kappa) \end{equation} with 
$D=ME$.  In the coarsening process the cells move in a size space $\{a\}$ 
(some phase space) with a velocity (1).  We can write a continuity 
equation in this space as \begin{equation} \partial f/\partial t+\partial 
(f{\rm d}a/{\rm d}t)/\partial a=0 \end{equation} for a cell size 
distribution function $f(a,t)$ \cite{6,7,8}.  The function $f(a,t)$ 
is normalised so that $n(t)=\int_{0}^{°}f{\rm d}a$ is the mean number of 
cells per unit volume; \begin{equation} \Omega_d\int_{0}^{°}a^{d}f{\rm 
d}a=1; \end{equation} $\bar{a}(t)=\int_{0}^{°}af{\rm d}a/n$ is the mean 
cell size and $\rho(t)=\Sigma/V=(\Omega_d d/2)\int_{0}^{°}a^{d-1}f{\rm 
d}a$ is the mean cell interface density.  Here $d=2,3$ is the cellular 
structure dimensionality and $\Omega_d=2\pi(d-1)/d$.  Condition (3) is 
nothing that the conservation law of the total volume of all cells in unit 
volume.  A remarkable property of this statistical dynamic system, 
discovered by LSH, is the occurrence of a {\it stable steady-state regime 
} of evolution.  This regime is observed if and only if the mean curvature 
$\kappa(t)$ take the unique "self-consistent" value 
$\kappa(t)=\sqrt{2/(Dt)}$.  All other macroscopic system variables obey 
also simple {\it scaling laws}:$\bar{a}(t) \propto \rho^{-1}(t) \propto 
t^{1/2}$, $n(t) \propto t^{d/2}$ and the solution of (1)-(3) has a {\it 
universal scaling form} \begin{equation} f_{0}(a,t) \propto 
(1/\bar{a}(t))^{d+1}\varphi(\mu_{1}a/\bar{a}(t)), \end{equation} where 
$\varphi(z)$ is a peaked function, depending on $d$ only, and 
$\mu_{1}\approx1$ \cite{6,7,8}.  
  \par There is a great number of publications 
devoted to the question of "improvement" of the individual cell dynamics 
equation (1) (we will say microscopic dynamics equation) to get a good 
agreement of the cell size distribution $f_{0}(a,t)$ with experimental 
data, essentially in the metallurgic literature related to normal grain 
growth \cite{2,3,4,9,10}.  Surprisingly small attention has been paid to 
mesoscopic and macroscopic behaviours of the cellular structures as some 
continuous elastic cellular medium with viscous motion law, although the 
existence of interesting self-organisation effects is almost evident here 
\cite{11,12}.  Indeed, if we have a good look at Fig.  1 we can see a 
strong size correlation of adjacent cells.  Near a small cell the 
probability to find another small cell is grater then a large one, and 
vice versa, near a large cell the probability to find another large cell 
is grater then a small one.  In other words we observe here 
space-correlated fluctuations of the mesoscopic cell interface density 
$\rho({\bf x},t)$.  Qualitatively this effect can be easy analysed.  Let 
us consider a nonhomogeneous cellular structure with the macroscopic mean 
interface density $\rho(t)$ and the mesoscopic one $\rho({\bf 
x},t)=\rho(t)+\delta \rho({\bf x},t)$ , weakly varying in space.  To be 
more precise, let us assume that $\delta \rho({\bf x},t)<<\rho(t)$ and the 
wavelength $\lambda$ of spatial variations of $\rho({\bf x},t)$ is much 
greater than $\bar {a}\approx1/\rho(t)$.  Let us cut out a medium volume 
$V_L$ having the size $L<<\lambda$, but still including many cells.  Each 
element ${\rm d}\bf s$ of the surface $\Sigma_L$, enclosing $V_L$, is 
affected by the interface tension of the cells, surrounding $V_L$.  The 
total interface tension force acting on $V_L$ is $\oint E \rho {\rm d}{\bf 
s}=\int E \nabla \rho {\rm d}V$.  In a homogeneous cell medium this force 
equals zero.  In the nonhomogeneous medium the force is non zero.  The 
total cell interface in $V_L$, which moves as the whole under the force 
effect, is $\int \rho{\rm d} V$.  Hence the force acting on cell interface 
unit is $E\nabla\rho/\rho$.  Supposing the viscous law for the cell 
interface movement, we find a cell interface flux ${\bf j}=\rho{\bf 
v}\propto D \nabla \rho$.  We see that the cells make a collective motion 
in the direction $\nabla\rho$.  This is an Òuphill diffusion" of the cell 
medium.  Taking into account the annihilation law ${\rm d}\rho/{\rm 
t}\propto-D \rho^3$, which follows from $\rho(t)\propto t^{-1/2}$, we get 
\cite{12} \begin{equation} \partial \rho/\partial t=-\gamma_1 D 
\rho^3-\gamma_2 D \triangle \rho.  \end{equation} Here $\gamma_1$ and 
$\gamma_2$ are dimensionless constants, depending on the scaling 
distribution function $\varphi(z)$, only.  Equation (5) has the form of a 
diffusion-reaction equation with the cell interface density $\rho$ playing 
the role of concentration and with a negative "diffusion" coefficient.  We 
see that, the amplitude of any small spatial fluctuation $\delta \rho({\bf 
x},t)$ with wavelength $\lambda<\xi$, where $\xi=2\pi \sqrt{\gamma_2 / 3 
\gamma_1}\rho^{-1}$ is the correlation length of the fluctuations, grows 
without bound.  It follows that competing stochastic processes of cell 
disappearance and cell medium "uphill diffusion" can give rise to the 
formation of large-scale domains of the size $\lambda \leq \xi$ having the 
density $\rho({\bf x},t)$ greater or smaller than the mean value $\rho(t)$ 
(Fig.  2 \cite{13}).
\begin{figure}
\includegraphics{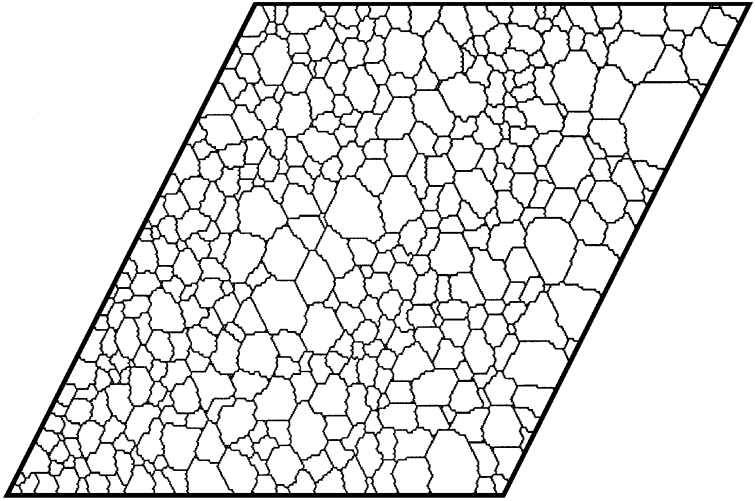}
\caption{Late-time configurations in Potts model of grain growth 
simulation.}
\end{figure}
  It also means that space derivatives of higher order of $\rho$ 
must be taken into account in right side of (5) for detailed analysis of 
the system stability and more accurate estimation of $\xi$.  \par Whereas 
the physical ideas explaining the possible macroscopic self-orga\-ni\-sa\-tion 
effects are simple, the consecutive derivation of the mesoscopic evolution 
equation (5) and the calculation of kinetic coefficients $\gamma_1$ and 
$\gamma_2$, defining the effect quantitatively, turns out to be a 
difficult statistical task \cite{14}.  Therefore the "technique" of the 
problem solution asks for some commentaries.  Let us consider an ideal gas 
as a well-known example for comparison.  The consistent description of its 
evolution include three classic scale levels of consideration: micro-, 
meso- and macroscopic \cite{5,6,15,16}.  On the {\it microscopic} level 
the evolution is effectively described by methods of physical kinetics, 
i.e.  by Boltzmann {\it kinetic equation} \cite{5,6} for the system 
probability distribution function $f$ in a phase space, taking into 
account interaction and movement laws of atoms.  Having the equilibrium 
solution $f_0$ for the homogeneous gas (Maxwell-Boltzmann (MB) 
distribution), it is possible to move to the {\it mesoscopic level} of the 
system description by introducing mesoscopic quantities 
(density, pressure, etc.) as moments of the distribution function.  In 
the case of small spatial gradients of the mesoscopic quantities, 
integration of kinetic 
equation by Enskog method gives a {\it hydrodynamic} equation 
\cite{6,15}.  If the hydrodynamic equation has steady-state nonhomogeneous 
solutions, it is possible to talk about a {\it macroscopic} description of 
the system by means of spatial patterns of size $\xi$ much greater than 
the mesoscopic size.  To study the patterns formation it is convenient to 
transform the hydrodynamic equation into Langevin {\it stochastic} 
equation \cite{16} for an {\it order parameter}.  As we 
see above LSH distribution (4) plays the same outstanding role for the 
coarsening cellular structures in steady-state, as MB distribution plays 
for ideal gases in equilibrium state \cite{5,6,7,8}.  All difference 
between MB and LSH distributions is that the first is determined by 
thermofluctuations related to collisions of gas atoms and the second is 
determined by stochastofluctuations related to the process of cells 
annihilation.  Hence the main methodical idea of this paper is to use LSH 
distribution as a basis for the multilevel description of cellular 
structure, like MB distribution is used 
in order to go from the kinetic equation to the hydrodynamic one and then 
to the stochastic one.  This idea has interesting applications for other 
physical systems, where the scaling regime is reached.  For example, its 
application allows to generalise Lifshitz-Slyozov theory of homogeneous 
coalescence \cite{7} for the nonhomogeneous case \cite{17}, to describe 
the relaxation of nonhomogeneous dislocation structure during annealing 
\cite{12} etc.

\section{Kinetic theory (microscopic level)}

 Let us consider as in Sec.  1 
a nonhomogeneous cellular structure with the interface density $\rho ({\bf 
x},t)$, weakly varying in space.  We divide the volume $V$ of the system 
into small coarse-graining elements $V_L$ with $L<<\lambda$, but still 
including many cells.  In the first approximation the cell dynamics of the 
structure in each $V_L$ will be defined by the dynamics equation (1), with 
the curvature $\kappa$ becoming a local caracteristic - an {\it interface 
curvature self-consistent mean field} $\kappa({\bf x},t)$ .  The field 
$\kappa({\bf x},t)$ in $V_L$ at point {\bf x} equals to the 
self-consistent mean of cell curvature $\kappa (t)$ determined by (1)-(4) 
for the cell structure in $V_L$ if we approximately consider it as 
homogeneous.  The field $\kappa({\bf x},t)$ is a weakly varying function 
in space as $\rho ({\bf x},t)$ is.  In the second approximation the 
dynamics equation (1) will contain gradients of the field $\kappa({\bf 
x},t)$.  It is easy to establish this fact if we analyze an 
one-dimensional cellular structure ($d=1$).  Let us cell interfaces are 
located at the points $x_{i} (i=0,\pm1,\pm2,\ldots)$ of $x$-axe.  We have 
the cell radii $a^{(i)}=(x_{i+1}-x_{i})/2$ , the cell curvatures 
$\kappa^{(i)}=1/a^{(i)}$ and the cell centres coordinates 
$x^{(i)}=(x_{i+1}+x_{i})/2$.  Each interface is acted upon by attraction 
of adjacent ones (the interface tension of the cells) according to the law 
$F_{i}=E(\kappa^{(i)}-\kappa^{(i-1)})$ with the result that the cell 
changes its position $dx^{(i)}/dt=D(\kappa^{(i+1)}-\kappa^{(i-1)})/2$ and 
its size $da^{(i)}/dt=D(\kappa^{(i+1)}+\kappa^{(i-1)}-2\kappa^{(i)})/2$.  
If we now introduce a curvature exact field $\kappa_{e}({x},t)$ as a step 
function, which is $\kappa^{(i)}$ if $x_{i}<x<x_{i+1}$ , we can rewrite 
$dx^{(i)}/dt=D[\kappa_{e}(x^{(i)}-a^{(i)},t)-\kappa_{e}(x^{(i)}+a^{(i)},t)]/2$ 
and 
$da^{(i)}/dt=D[\kappa_{e}(x^{(i)}+a^{(i)},t)+\kappa_{e}(x^{(i)}-a^{(i)},t)-2/a^{(i)}]/2$.  
Replacing the exact field $\kappa_{e}({x},t)$ by the mean field 
$\kappa({x},t)$ and taking into account that $\bar{a}<<L<<\lambda$ we get 
respectively the cell velocity and the cell size change rate 
\begin{equation} {\rm d}{\bf x}/{\rm d}t =D a \nabla \kappa/d, 
\end{equation} \begin{equation} {\rm d}a/{\rm d}t =-D (1/a-\kappa)+Da^2 
\triangle \kappa/(2d).  \end{equation} It is easy to understand that the 
abovementioned definition of the curvature mean field $\kappa_{e}({\bf x},t)$ 
holds for nonhomogeneous cellular structures in two ($d=2$) and three 
dimensions ( $d=3$).  Expressions (6) and (7) are also valid.  Really, 
each cell of the structure may be approximately thought of as having the 
form of sphere.  Then the all cellular structure configuration may be 
described through the cell centre coordinates and the cell radii.  Let us 
choose a cell of radius $a$ with centre in a point {\bf x} and an 
interface point with coordinate ${\bf x'}={\bf x}+{\bf a}({\bf x'},t)$ 
of the cell.  
The velocity ${\bf w}({\bf x'},t)$ of the interface point affected by the 
interface tension force ${\bf F}({\bf x'},t)=-E[1/a(t)-\kappa_{e}({\bf 
x'},t)]{\bf a}({\bf x'},t)/a(t)$, is determined by the viscous law ${\bf 
w}({\bf x'},t)=M{\bf F}({\bf x'},t)]$.  Replacing $\kappa_{e}({\bf x},t)$ 
by $\kappa({\bf x},t)$ we can approximately calculate the cell velocity 
${\rm d}{\bf x}/{\rm d}t =\oint w({\bf x'},t) {\rm d} {\bf s'}/ (\Omega_d 
da^{d-1})\approx a \nabla w/d$ and the cell size change rate ${\rm d} a 
/{\rm d}t =\oint w({\bf x'},t) {\rm d} s'/ (\Omega_d da^{d-1})\approx w+ 
a^2 \triangle w /(2d)$ and find expresions (6) and (7).  
   \par We see that the 
cells move in a {\it coordinate-size space} $\{ {\bf x},a\}$ in the 
process of coarsening.  These movements are conveniently described by 
kinetic equation \cite{18} \begin{equation} \partial f/ \partial 
t+\partial (f{\rm d}{\bf x}/{\rm d}t)/\partial {\bf x}+\partial (f{\rm 
d}a/{\rm d}t)/\partial a=0 \end{equation} for the structure probability 
distribution function $f({\bf x},a,t)$ in the cell coor\-di\-nate-size space 
with normalisation condition \begin{equation} (\Omega_d / V)\int {\rm d} 
{\bf x} \int_{0}^{°}{\rm d}a a^{d}f=1.  \end{equation} Equations (8) and 
(9) have the physical sense of continuity equation in the coordinate-size 
space and of conservation law of the total volume of all cells of the 
system.  They are natural generalisations of (2) and (3).  For the 
macroscopically homogeneous cell structure the set of equations (6)-(9) 
comes to (1)-(3) and has the solution (4) with $\varphi 
(z)=[dz/(2-z)^{d+2} ](2e)^d exp[2d/(z-2)]$ for $0<z<2$ and $\varphi (z)=0$ 
for $z\geq2$ ; $\mu_i=\int_{0}^{°} z^{i}\varphi (z){\rm d}z$ .  The mean 
structure variables obey simple scaling laws: $\bar{a}(t)=\mu_1 
\sqrt{Dt};$ $\kappa (t)=(\mu_{d-2}/\mu_{d-1})/\sqrt{Dt};$ 
$n(t)=1/(\Omega_d \mu_d (\sqrt{Dt})^d)$ and $\rho(t)=d \mu_{d-1}/(2\mu_d 
\sqrt{Dt})$.  
   \par To find $f_1({\bf x},a,t)=f_0 (a,t)+\delta f({\bf x},a,t)$ 
in the first-order approximation for small parameters $\mid \nabla \rho 
\mid / \rho^2 <<1$, it is necessary, as in well-known Enskog method 
\cite{6}, to substitute $f_1({\bf x},a,t)$ in (6)-(9) and linearize them, 
taking into account that $\delta f({\bf x},a,t)<<f_0 (a,t)$.  I hope in a 
future work to realise this possibility, but this is proving at present to 
be a difficult task \cite{14}.

\section{Hydrodynamic theory (mesoscopic level)} 

Equation (8) provides the 
kinetic (microscopic) description of the cellular structure coarsening.  
To get the hydrodynamic equation let us rewrite (4) as \begin{equation} 
f_0 (a,t)= \alpha \rho^{d+1} \varphi (\beta a \rho ), \end{equation} where 
$\alpha=(\mu_d)^d /[\Omega_d (d \mu_{d-1})^{d+1}]$ and $\beta=\mu_d/(d 
\mu_{d-1})$ are constants, and use again as in Sec.  1 and 2 the standard 
coarse-grain averaging procedure \cite{1}.  Namely, let us introduce a 
mesoscopic coarse-graining length $L(t)>> \bar{a}(t)$, so that a 
mesoscopic coarse-graining volume $V_L =\Omega_d L^d$ includes many 
cells.  Let us divide the volume $V$ of cellular structure into elements 
of volume $V_L$.  For small gradients of mesoscopic variables $\kappa({\bf 
x},t),\bar{a} ({\bf x},t),\rho ({\bf x},t),$ and $ n({\bf x},t)$ we 
may suppose that a {\it local homogeneous steady-state} is reached in each 
separate element $V_L$ after a time $\tau_L =L^2 / D$ much greater than 
the cell annihilation time $\tau_a =\bar{a}^2 / D$.  For all this, the 
cellular structure as the whole (macroscopically) is not in the 
homogeneous state.  Then the distribution function $f_0 ({\bf x},a,t)$ in 
$V_L$ at point $\bf x$ can be assumed to be a {\it local steady-state 
function}, equal to LSH distribution (10) for the homogeneous cellular 
structure with density $\rho ({\bf x},t)$, that prevails at the point $\bf 
x$ .  This supposition is completely analogous to the {\it local system 
equilibrium} one in {\it linear thermodynamics} \cite{5,6}.  Multiplying 
(8) by $\Omega_d d a^{d-1}/2$, (6) by $\Omega_d d a^{d-1} f_0 /2$, (8) by 
$\Omega_d a^d$ , integrating them by $a$ and using conservation law (9), 
we make the averaging procedure.  After sufficiently heavy calculations we 
obtain, respectively, the {\it equation of cell medium continuity} , an 
analogue of {\it Darcy's law} in hydrodynamics and an equation similar to 
the {\it state equation} in gas dynamics \cite{15}: \begin{equation} 
\partial \rho/ \partial t+\nabla (\rho {\bf v} ) =-D[ \alpha_1 \rho^3- 
\alpha_2 \rho^2 \kappa -\alpha_3 \triangle \kappa] ,\end{equation} 
\begin{equation} {\bf v}=(D/2) \rho^{-1} \nabla \kappa ,\end{equation} 
\begin{equation} \rho^3 \kappa = \beta_1 \rho^4 - \beta_2 \rho \triangle 
\kappa + \beta_3 \nabla \kappa \nabla \rho. \end{equation} The set of 
three hydrodynamic equations (11)-(13), containing three hydrodynamic 
quantities of the
 cell medium: the density $\rho ({\bf x},t)$ , the velocity ${\bf 
v}({\bf x},t)=(\Omega_d d / \rho) \int_{0}^{°} a^{d-1} ({\rm d}{\bf 
x}/{\rm d}t )f {\rm d} a $, and the curvature $\kappa ({\bf x},t)$, forms 
a closed set.  The role of the quantities $\rho ({\bf x},t)$ , 
$\kappa ({\bf x},t)$ and ${\bf v}({\bf x},t)$ are completely analogous to the 
role of gas density, pressure and velocity in viscous gas dynamics.  If 
initial and boundary conditions are known, the cell medium movement can be 
determined by traditional hydro-gas dynamics methods \cite{15} for any 
continuous distribution of $\rho ({\bf x},t)$ , 
$\kappa ({\bf x},t)$ and ${\bf v}({\bf x},t)$.  
By substituting (12)-(13) in (11), the set can be 
reduced to evolution equation (5).  In (5), (11)-(13) $\alpha_1, 
\alpha_2,$ $\beta_1, \beta_2,$ $\gamma_1,\gamma_2$ are calculated 
constants.  For $d=2$: $\alpha_1=1.54, \alpha_2=1.11, 
\alpha_3=0.25,$ $ \beta_1=1.11, \beta_2=0, \beta_3=0.267,$ 
$\gamma_1=0.308, \gamma_2=0.277 $.  For $d=3$: 
$\alpha_1=1.16, \alpha_2=1.44, \alpha_3=0.5,$ $ \beta_1=0.718, 
\beta_2=0.132, \beta_3=0.263,$ $ \gamma_1=0.129, \gamma_2=0.136$.  
Full expressions are sufficiently bulky (e.g.  
$\alpha_1= 4(d-1) \mu_d ^2 \mu_{d-3} /( d^2 \mu_{d-1}^3)$) and will be 
published elsewhere \cite{14}.

\section{Stochastodynamic theory (macroscopic level)} 

To analyse the 
dynamics of macroscopic spatial perturbations of the cell me\-dium, let us 
introduce a relative value of deviation of the cell interface density from 
its mean macroscopic value $\psi ({\bf x},t)= \delta \rho ({\bf x},t) / 
\rho (t) <<1$ , where $\rho (t) $ is the uniform solution of (5).  
Substituting $\rho ({\bf x},t)=\rho (t)+\delta \rho ({\bf x},t)$ in (5) 
and coming to new time and space variables 
$\partial \tilde {t}=3\gamma_1 D \rho^2 (t)\partial{t}=-\partial \ln \rho^3 (t)$ 
and $\partial{\tilde {\bf x}}=\rho(t)\partial{\bf x}$ 
(intrinsic system variables \cite{7}), we obtain \begin{equation} 
\partial \psi / \partial \tilde {t}=- \delta {\cal F}/ \delta \psi  
\end{equation} with 
${\cal F}\{ \psi \}=\int {\rm d} \tilde {\bf x} [ {\cal A} \psi^2 + 
{\cal C} (\tilde {\nabla} \psi )^2]/2$ , ${\cal A}=1$ and ${\cal 
C}=-\gamma_2 / (3 \gamma_1 )$ (${\cal C}=-0.300$ for $d=2$ and ${\cal 
C}=-0.351$ for $d=3$ ).  Taking $\psi (\tilde{{\bf x}}, \tilde{t})$ in the 
form $\psi (\tilde{{\bf x}}, \tilde{t})=\psi (\tilde{{\bf k}}, 
\tilde{\omega}) e^{i(\tilde{{\bf k}}\tilde{{\bf x}}-\tilde{t} 
\tilde{\omega})} $, we get the dispersion relation 
$\tilde{\omega}(\tilde{{\bf k}})=-i \tilde{\sigma}(\tilde{{\bf k}})$ , 
where \begin{equation} \tilde{\sigma}(\tilde{{\bf k}})={\cal A}+{\cal C} 
\tilde {k} ^2 \end{equation} is the damping coefficient.  Including in 
${\cal F}\{ \psi \}$ the term $-h\psi$ describing the effect of an external field 
$h (\tilde{{\bf x}}, \tilde{t}) \propto e^{i(\tilde{{\bf k}}\tilde{{\bf 
x}}-\tilde{t} \tilde{\omega})} $ , we find the generalised susceptibility 
$\tilde{\chi}(\tilde{{\bf k}}, \tilde{\omega}) =1/ 
[\tilde{\sigma}(\tilde{{\bf k}})-i \tilde{\omega}]$.  
   \par As all mesoscopic 
quantities, $\psi$ is a fluctuating quantity.  But on writing (14) we have 
neglected these fluctuations.  The quantity $\psi $ is a sum of a number 
of microscopic random quantities.  Each of them corresponds to the 
contribution of a cell being in $V_L$.  In view of central limit theorem 
of statistics, $\psi $ has Gaussian distribution function with the mean 
value $<\psi >_L =0$ and the dispersion $<\psi^2 >_L =\nu/[V_L 
N]$ , where $\nu=\mu_{2(d-1)}/ \mu^2_{d-1}-1$ ($\nu=0.109$ for 
$d=2$ and $\nu=0.373$ for $d=3$) \cite{14}.  Equation (14), after taking 
into account the fluctuations, becomes Langevin stochastic equation with 
${\cal F}\{ \psi \}$ being a {\it stochastodynamic potential} \cite{19}.  The 
noise $\zeta$ is Gaussian noise, with $<\zeta>=0$ and the correlation 
function $<\zeta(\tilde{{\bf x}}, \tilde{t}) \zeta(\tilde{{\bf x}}', 
\tilde{t}')>=2\theta \delta (\tilde{{\bf x}}-\tilde{{\bf x}}') 
\delta(\tilde{t}-\tilde{t}')$ , defined by the fluctuations at scales 
$a<<L$ and $\tau_a<<\tau_L$.  Here $\theta= \Omega_d(d/2)^{d-1}( \mu_{d-1} 
^d / \mu_{d} ^{d-1})(\mu_{2(d-1)} / \mu_{d-1}^2-1)$ ($\theta=0.619$ for 
$d=2$ and $\theta=8.09$ for $d=3$ ) \cite{14}.  The forms of evolution 
equation (14) and of the noise $\zeta$ guarantee that the steady-state 
probability distribution functional is given by Boltzmann formula $P\{ 
\psi \}\propto exp[-{\cal F}\{ \psi \}/\theta]$ \cite{19}.  We see that our cell 
system is rated in the class of so-called potential systems with potential 
${\cal F}\{ \psi \}$ having Landau-Ginzburg form \cite{5,6,16} and the quantity 
$\psi$ plays a role of an {\it order parameter}.  It is not surprizing because 
in general sense the system is in perfect analogy to a thermodynamic 
system with a nonconserved order parameter $\psi$.  So the 
forms of evolution equation (14) and of potential ${\cal F}\{ \psi \}$ are 
dictated by simple symmetry macroscopic considerations.  Instead of the 
thermal noise in thermodynamic systems we have here the stochastic noise, 
created by the continuous process of cells annihilation.  In virtue of 
reasoning, fluctuation-dissipation theorem may be applied, according to 
which the generalised susceptibility $\tilde{\chi}(\tilde{{\bf k}}, 
\tilde{\omega})$ determines the spectral correlation function of 
fluctuations as $\tilde{S}(\tilde{{\bf k}}, \tilde{\omega})=(2 \theta/ 
\tilde{\omega}) {\rm Im}\tilde{\chi}(\tilde{{\bf k}}, \tilde{\omega})$ .  The 
structure function has usual Ornstein-Zernike form $\tilde{S}(\tilde{{\bf 
k}})=\theta/ \tilde{\sigma}(\tilde{{\bf k}})$ \cite{5,6}.  
   \par From (15) we 
see, that the order parameter modes $\psi (\tilde{{\bf k}}, \tilde{t} )$ 
with $\tilde {k} > \sqrt { - {\cal A} / {\cal C} }$ are unstable.  This is 
every indication that highest order terms of the expansion of ${\cal F}\{ \psi 
\}$ in powers of $\tilde{\nabla} \psi$ must be considered for detailed 
analysis of the system stability.  The derivation of the exact expression 
for these terms by Enskog method is a very heavy task \cite{6,14} not 
realized to the present.  We proceed now differently.  Let us use the fact 
that our system is potential with the nonconserved order parameter 
$ \psi$ .  Then we may consider ${\cal F}\{ \psi \}$ as some 
phenomenological potential in the abovemensioned Landau sense.  This 
potential have necessarily the general form $ {\cal F}\{ \psi \}=\int {\rm d}{\bf 
x} [ {\cal A} \psi^2 + {\cal C} (\tilde {\nabla} \psi )^2+{\cal D} (\tilde 
{\triangle} \psi )^2]/2 $ with ${\cal D}>0$ \cite{5,16,20}.  The 
phenomenological coefficients ${\cal A}$,${\cal C}$ and ${\cal D}$ can be 
found from experiments, computer simulations or by a direct calculation as 
we have made above for ${\cal A}$ and ${\cal C}$ .  As a consequence we 
get the damping coefficient in the form \begin{equation} 
\tilde{\sigma}(\tilde{{\bf k}})={\cal A}+{\cal C} \tilde {k} ^2+{\cal 
D}\tilde {k} ^4 .\end{equation} We do not know the real value of ${\cal D}$ 
, but we are able to forecast the possible scenarios of system evolution 
\cite{20}.  If ${\cal A}>0$, ${\cal C}<0$ and ${\cal D}>{\cal C} ^2 
/(4{\cal A}) $ the evolution equation (14) has the single stable solution 
$\psi_0 (\tilde{{\bf x}})=0 $ , and modes $\psi (\tilde{{\bf k}}, 
\tilde{t})$ with any $\tilde{{\bf k}}$ decay with time.  Therefore the 
system is stable with respect to large-scale fluctuations of $\psi 
(\tilde{{\bf x}}, \tilde{t}) $ near $\psi_0 (\tilde{{\bf x}})=0 $ .  The 
modes $\psi (\tilde{{\bf k}}, \tilde{t}) $ with $\tilde{ k} =\tilde{ k}_{ 
\cal C} =\sqrt{-{ \cal C}/ (2 {\cal D })}=2 \pi/ \tilde{\xi}$ , where 
$\tilde{\xi}$ is an intrinsic correlation length, have the lowest decay 
rate.  That is why under the action of the stochastic internal noise 
$\zeta$ , created by a continuous process of cells annihilation, 
space-correlated fluctuations arise in the system.  The system is in a 
{\it modulated state} (short range ordering) and the structure function 
\begin{equation} \tilde{S}(\tilde{{\bf k}})= \theta/ ({\cal A}+{\cal C} 
\tilde {k} ^2+{\cal D}\tilde {k} ^4 ).  \end{equation} has a maximum at 
$\tilde{ k}=\tilde{ k}_{\cal C}$ (Fig.  3), reflecting the abovestated 
fact that the modes $\psi (\tilde{{\bf k}_{\cal C}}, \tilde{t}) $ decay with the 
lowest rate.  
\begin{figure}
\includegraphics{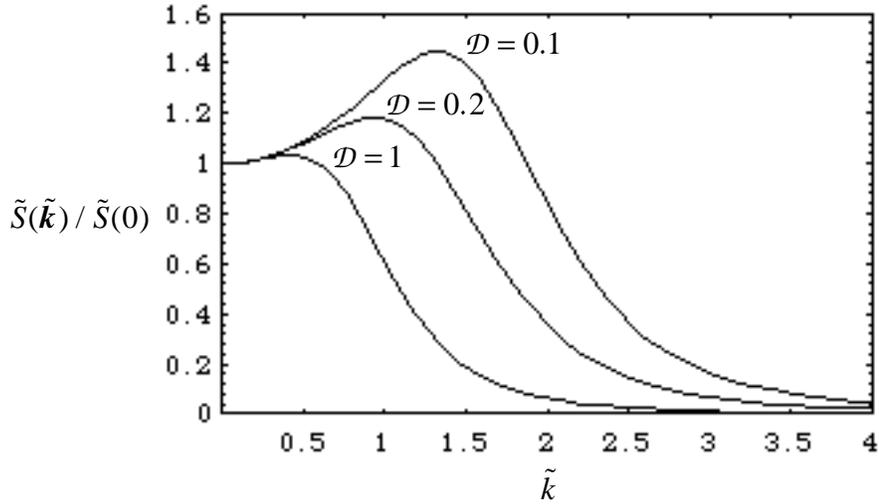}
\caption{Scaled structure function of fluctuations 
$\tilde{S}(\tilde{{\bf k}})/\tilde{S}(0)= 1/ [1+
({\cal C}/{\cal A}) \tilde {k} ^2+({\cal D}/{\cal A})\tilde {k} ^4 ] $ 
of the relative interface density 
$\psi ({\bf x},t) =  \rho ({\bf x},t) / \rho (t) -1 $ 
for different ${\cal D}$.}
\end{figure}
If ${\cal A}>0$, ${\cal C}<0$ and $0<{\cal D}<{\cal C} ^2 
/(4{\cal A}) $ , the evolution equation (14) has the unstable solution 
$\psi_0 (\tilde{{\bf x}})=0 $ and stationary one-, two-, ...  $d$ 
-dimensional space-periodical solutions with wavelength 
$\tilde{\lambda}=\tilde{\xi}$ \cite{20}.  Under the action of the 
stochastic internal noise $\zeta$ the system comes from the homogeneous 
state into one of these {\it patterned state} (long range ordering).  The 
question of existence of the first or of the second type of ordering 
depends of the value of coefficient ${\cal D}$ .

\section{Discussion}

 We see that in the process of structure coarsening 
the macroscopic cell interface density $\rho (t)$ decreases with time as 
$\rho (t)\propto t^{-1/2}$ and the mesoscopic density $\rho ({\bf x},t)= 
\rho(t)[1+\delta \rho ({\bf x},t) / \rho (t)]$ also decreases (Fig.  1).  
The competition of two stochastic processes of the cells annihilation and 
of the cells collective motion creates space-correlated fluctuations of 
the cell interface density.  These fluctuations are the adjacent domains 
of size $\xi$ with $\rho ({\bf x},t)$ greater or smaller than $\rho (t)$ 
(Fig.  1,2).  In other words, they are the domains in which the cells have 
the mean size ${\bar a} ({\bf x},t)$ greater or smaller than the 
macroscopic mean value ${\bar a} (t)$ .  The amplitude of relative 
fluctuations $\delta \rho ({\bf x},t) / \rho (t)$ and the domain size 
(characteristic wavelength of fluctuations), divided by the mean cell size 
$\xi / {\bar a} (t) \propto \xi \rho (t)=\tilde{\xi}$ stay constant.  
These domains can be short-range ordered (Fig.  1) or long-range (Fig.  2) 
ordered in space.  The short-range ordering of cell interface density 
fluctuations resemble the fluid density fluctuations ordering in the 
well-known phenomenon of critical opalescence.  The long-range ordering of 
cell interface density fluctuations resemble to the effect of spinodal 
decomposition in thermodynamic systems.  
   \par This is just a moment to return 
to the question discussed in Sec.  1 and 2 about basing of the 
self-consistent mean field approximation for description of the cellular 
systems.  The condition of its validity is the condition that the relative 
value of field fluctuations $\delta \kappa ({\bf x},t) / \kappa (t) 
\approx \delta \rho ({\bf x},t) / \rho (t)\approx \psi({\bf x},t)$ is 
small.  More priscisely, the fluctuations averaged over the 
coarse-graining volume $V_L$ must be small:$<\psi^2 ({\bf x}, t)>_L 
=\nu/[V_L N({\bf x}, t)]<<1$ .  Choosing the volume of a Òfirst 
coordination sphereÓ as a minimum possible value of $V_L$ and taking into 
consideration that $n=V_L N$ is the number of cells in $V_L$ , we see that 
this condition is well satisfied both for $d=2$($\nu=0.109, n \approx 7$) 
and $d=3$($\nu=0.373, n \approx 13$).  The general condition of validity 
of mean-field theory for the systems with Landau-Ginzburg potential ${\cal F}\{ 
\psi \}$ is well-known Ginzburg criterion: fluctuations of $\psi$ averaged 
over the correlation volume must be small $<\psi^2 ({\bf x}, t)>_\xi 
=\nu/[V_\xi N({\bf x}, t)]<<1$ .  This criterion is obviously fulfilled if 
$L\le \xi$ .  Uniting this condition with the foregoing condition of 
existence of stochastodynamic potential ${\cal F}\{ \psi \}$ , we have finally 
${\bar a}<<L\le \xi$ .  This is just the familiar condition of the choice 
of the coarse graining size $L$ , which is also the condition of validity 
of the self-consistent mean field approximation in phase transformation 
theory.  It can be satisfied if ${\bar a}<<L\le \xi$ .  In coarsening 
cellular systems this is always the case for the short range ordering 
(${\cal D}>{\cal C} ^2 /(4{\cal A}) $) and this is so indeed if ${\cal 
D}>>-{\cal C} /(8\pi^2) $ for the long-range ordering ($0<{\cal D}<{\cal 
C} ^2 /(4{\cal A}) $ ).  
   \par In spite of that the ordering effects of cell 
interface density fluctuations can be evidently established by visual 
analysis of different cellular structures in experiments (Fig.  1) and 
computer simulations (Fig.  2), I am not aware of publications, where the 
structure function (17), describing the effect quantitatively, has been 
directly obtained.  I feel that it is related to two reasons.  Firstly, it 
must be noted that hydrodynamic and stochastodynamic theories developed 
here are valid in the long wavelength limit $\tilde{k}=k \rho (t)<1$ , 
when we can guarantee the possibility of application of the cell medium 
conception and the self-consistent mean field approximation.  In the long 
wavelength limit $\tilde{k}<1$ the maximum of $\tilde{S}(\tilde{{\bf k}})$ 
is hardly observed: if $\tilde{k}_{\cal C}<<1$ the maximum is small (Fig.  
3); if $\tilde{k}_{\cal C}\approx 1$ it is hardly separated from the other 
maximum $\tilde{k}_{0} \approx 2$ corresponding to the trivial effect that the 
mean distance between neighbouring interfaces is $2{\bar a}$ .  This later 
maximum cannot be described within our long wavelength approach and is not 
shown in Fig.  3.  Therefore the more strong confirmation of existence of 
the large-scale fluctuations of cell interface density is the observation 
of the {\it negative curvature} of $\tilde{S}(\tilde{{\bf k}})$ in the 
macroscopic $\tilde{k}\rightarrow 0$ limit (Fig.  3).  However it demands of 
a laborious statistical analysis or a scattering study of very large 
cellular systems.  In actual practice the finite-size induced effects 
stays too essential \cite{21}.  Secondly, in computer experiments, using 
usually " $q$- states Potts model" \cite{10,13}, the structure function 
$S({\bf k})$ is calculated as a Fourier transform of the correlation 
function of a phases volume density.  If all phases have the same 
equilibrium energy, and the boundaries separating the phase domains, are 
sufficiently distinct, the system state is completely defined by the total 
interfacial area.  In consequence, the cell interface density $\rho ({\bf 
x},t)$ and the corresponding structure function (17) is a much more 
sensible characteristic of the system than the phases volume density and 
the corresponding to it structure function.  Quite recently \cite{22} it 
has been demonstrated by computer simulation of normal grain growth that a 
size correlation between neighbouring grains exist in grain structure.  
Unfortunately the spectral correlation analysis of the cell interface 
density fluctuations by means of the structure function (17) has not been 
made and direct quantitative comparison of these results with present 
theory results is difficult.  Still there is no doubt that we are dealing 
with the first observation of the short-range ordering of cell interface 
density fluctuations in cellular structures.  New experiments or computer 
simulations with careful space spectral analysis of cell interface density 
fluctuations are needed to confirm the existence of dynamic 
patterning in cellular structures.

\section{Summary} 

In summary, the problems of multilevel description of 
cellular structure evolution are considered from positions of kinetic, 
hydrodynamic and stochastodynamic theories of nonequilibrium statistical 
systems.  It is demonstrated that the effective method to solve the 
problem is to describe the cellular structure on the microscopic level by 
the probability distribution function in the coordinate-size space, i.e.  
in the space of the cell coordinates and the cell sizes.  The kinetic 
equation for the weakly nonhomogeneous cellular structure is written and 
the development to hydrodynamic equations for a   
continuous elastic cellular medium with viscous motion law is realised.  
These non-linear differential equations contain three mesoscopic 
quantities of the cell medium (the cell interface density, the cell medium 
velocity and the cell interface curvature) and are analogous to the 
continuity equation, Darcy's law and the equation of state in gas 
dynamics.  For continuous flow of the cell medium the hydrodynamic 
equation set is reduced to a single evolution equation for the cell 
interface density.  It has the form of a diffusion-reaction equation with 
a negative "diffusion" coefficient and with the cell interface density 
playing the role of "concentration".  It is shown that the relative value 
of deviation of the cell interface density from its mean macroscopic value 
is a suitable order parameter of the system.  The order parameter dynamics 
are described by Langevin stochastic equation with a stochastodynamic 
potential having Ginzburg-Landau form and with a stochastic noise, created 
by the continuous process of cells annihilation.  Analysis of the order 
parameter dynamics in coarsening reveals a new effect of macroscopic 
patterning in the cell medium: competition between two stochastic 
processes of the small cells "eating up" by large ones and of the "uphill 
cells diffusion" leads to the creation of space-correlated fluctuations of 
the cell interface density.

\section{Acknowledgements}

I always keep in my memory stimulating discussions with the late B.  M.  
Strunin and I.  M.  Lifshitz carried out during the early stages of this 
work at the Moscow Engineering Physics Institute.  I acknowledge valuable 
discussions with V.  V.  Slyozov, A.  I.  Ryazanov, L.  P.  Kubin, V.  
Pontikis, N.  Rivier and R.  Lefever.  I am grateful to the Laboratoire
des Solides Irradi\'{e}s, Ecole Polytechnique, France for kind hospitality and 
support of this work.

\end{document}